\documentclass[aps,prl,twocolumn,floats,showpacs,superscriptaddress]{revtex4}
\usepackage{graphicx,epsfig}
\usepackage{times}
\usepackage{graphics,dcolumn,bm,float}
\usepackage{amssymb,amsmath,rotate,color}

\begin{document}
\unitlength 1 cm
\newcommand{\be}{\begin{equation}}
\newcommand{\ee}{\end{equation}}
\newcommand{\bearr}{\begin{eqnarray}}
\newcommand{\eearr}{\end{eqnarray}}
\newcommand{\nn}{\nonumber}
\newcommand{\vk}{\vec k}
\newcommand{\vp}{\vec p}
\newcommand{\vq}{\vec q}
\newcommand{\vkp}{\vec {k'}}
\newcommand{\vpp}{\vec {p'}}
\newcommand{\vqp}{\vec {q'}}
\newcommand{\bk}{{\bf k}}
\newcommand{\bp}{{\bf p}}
\newcommand{\bq}{{\bf q}}
\newcommand{\br}{{\bf r}}
\newcommand{\up}{\uparrow}
\newcommand{\down}{\downarrow}
\newcommand{\fns}{\footnotesize}
\newcommand{\ns}{\normalsize}
\newcommand{\cdag}{c^{\dagger}}

\definecolor{red}{rgb}{1.0,0.0,0.0}
\definecolor{green}{rgb}{0.0,1.0,0.0}
\definecolor{blue}{rgb}{0.0,0.0,1.0}

\title{Anderson Transition in Disordered  Bilayer Graphene }

\author{M. H. Zare}

\affiliation{Department of Physics, Isfahan University of
Technology, Isfahan 84156-83111, Iran}

\author{Mohsen Amini}

\affiliation{Department of Physics, Isfahan University of
Technology, Isfahan 84156-83111, Iran}

\author{Farhad Shahbazi}

\affiliation{Department of Physics, Isfahan University of
Technology, Isfahan 84156-83111, Iran}

\author{S. A. Jafari}

\affiliation{Department of Physics, Isfahan University of
Technology, Isfahan 84156-83111, Iran}
\affiliation{School of Physics, Institute for Research in Fundamental Sciences (IPM), Tehran 19395-5531, Iran}

\begin{abstract}
Employing the Kernel Polynomial method (KPM), we study  the
electronic properties of the graphene bilayers  in the presence of
diagonal disorder, within the tight-binding approximation. The KPM
method  enables us to calculate local density  of states
(LDOS) without need to exactly diagonalize the Hamiltonian. We
use the geometrical averaging of the LDOS's at different lattice
sites as a criterion to distinguish the localized states from
extended ones. We find that bilayer graphene undergoes Anderson
metal-insulator transition at a critical value of disorder strength.
\end{abstract}

\maketitle  
\pagebreak

\section {introduction}
Graphene (the 2D allotrope  of carbon with honeycomb structure)
has attracted tremendous attraction since its  isolation by
Novoselov {\it et al}~\cite{Novoselov2}.  Some of the features
that makes graphene distinct from previously known materials are
properties such as, possibility to control the charge carrier
types from hole to electron in the same sample by a gate
voltage,  high mobility of charge carriers (about two orders of
magnitude larger than the best silicon based semiconductors),
anomalous integer Quantum Hall effect (IQHE) at room temperature
~\cite{NetoRMP,QHE}. Among them, the high mobility of carriers
which is due to the vanishing of back-scatterings, makes graphene
favorable in fabrication of carbon-based electronic devices.

The  spectral and transport properties  of graphene are well
described within the tight-binding approximation. In this model
the two upper energy bands, which are due to the $\pi$ and
$\pi^{*}$ bonds of $p_{z}$ orbitals normal to the honeycomb lattice plane,
touch each other with a linear dispersion at the corners of the
Brillouin zone (the so called $K$-points)~\cite{Wallace}. In this picture
graphene is a semi-metal whose low energy excitations are
massless fermions propagating with Fermi velocity ($v_{F}$) of
about $1/300$ times the light speed. These excitations are called
Dirac-fermions, which, along with strong stability of graphene
structure (which is related to the in plane $\sigma$ bonds), are
responsible for the odd properties of graphene~\cite{NetoRMP}.

Employing graphene in electronic devices requires the opening of gap
in its electronic spectrum. This can be done by constraining its
geometry either to graphene nano-ribbons or quantum-dots. However
these methods affect the electronic transport because of
formation of rough edges and also enhancement of the Coulomb
blockade effects in the small size structures~\cite{Han}. Moreover,
by reducingthe size and symmetry, the selection rules
tend to get more restrictive, hence reducing the possible conduction channels. Nevertheless, it
has been shown that by applying an electric field normal to a graphene
{\em bilayer} a gap can be opened whose magnitude is proportional to the
intensity of the applied electric field ~\cite{McCann,McCann1}. The size of gap can be as large
as 0.1-0.3 eV, thereby making the bilayer graphene an attractive
candidate in carbon-based transistors ~\cite{Nilsson}.

The graphene bilayer consists of  two layers of carbon atoms in
honeycomb structure placed on top of each other, making the Bernal
stacking ~\cite{Nilsson1}. In this stacking the upper layer is
rotated 60 degree relative to the lower one, in such a way that
the atoms in one sublattice, {\it i.e} $A_{1}$ and $A_{2}$, are
on the top of each other while the atoms of the $B$-sublattice in
upper layer are placed on the top of the hallow center of hexagons
of the lower layer . In the absence of external perpendicular
electric field the band structure of graphene bilayer consists of
four bands, arising from the $\pi$-bonding between $p_{z}$
orbitals.  Two of these bands touch each other at zero energy
with a quadratic dispersion, and so give rise to the massive
quasi-particles also called chiral fermions~\cite{McCann}. The
other two bands are separated by energy scale corresponding to the
inter-plane hopping energy, $t_{\bot}$; one laying below the zero
energy and the other above it. The electronic structures of
bilayer as well as multilayer graphene have been extensively
investigated
recently~\cite{McCann,Nilsson,Berger1,Berger2,Nilsson2,Novoselov1,Koshino}.

 The single and double layer samples of graphene can be grown with
 high crystalline quality. However there are inevitable sources of
 disorder which may affect the electronic transport in the
 fabricated samples.  Typical forms of disorder includes surface ripples,
 topological defects, vacancies, ad-atoms, charge
impurities and polarization field of the substrate. Scaling
theory of localization predicts that all the electronic states
are localized in two-dimensional systems, once smallest amount of
disorder is introduced~\cite{AndersonLocalization}. According to
this theory, the single and bilayer of graphene are expected to
be insulators at zero temperature. However theoretical as well as
experimental studies show that  both single and bilayer graphene,
have a minimum conductivity of the order of conductance quantum
($e^{2}/h$) at the charge neutrality point ~\cite{Fradkin}. It is
also shown that the minimum conductivity  is at least twice
larger in the bilayer graphene ~\cite{Koshino,Snyman}. These
results are in contrast to prediction of the scaling theory of
localization and motivate us to study the localization properties
of electronic states in graphene systems. In our previous work,
we investigate the the single layer graphene in the presence of
on-site disorder by using the numerically powerful approach of
KPM~\cite{amini}. There, we found that disordered graphene
mono-layer shows Anderson transition at a critical value of
disorder intensity which is of the order of bandwidth. Our
calculation also brought an unusual theoretical prediction
according to which, the localization starts from the states near
the Dirac points and spreads toward the band edges by increasing
the amount of disorder~\cite{amini}. This type of metal-insulator
transition driven by short range diagonal disorder due to neutral
impurities, was experimentally verified in Hydrogen dozed graphene
by angular resolved photoemission experiment~\cite{Horn}.

In this paper we investigate the  localization properties of the
electronic states in bilayer graphene in presence of diagonal
disorder. We consider minimal coupling between two layers of
graphene, that is, we take into account only the hopping between
$A_{1}$ and $A_{2}$ sublattices. Here we use the kernel
polynomial method (KPM)~\cite{Weisse}, which consists in the
expansion of various spectral functions in terms of a complete
set of polynomials. We calculate different local density of
states (LDOS), from which we introduce a quantity for
distinguishing the localized states from extended ones. The CPU
time in this method grows proportional to square of the system
size, enabling us to study large lattice sizes in a moderate time.

\section { Model Hamiltonian}

As was mentioned before, we use a minimal  tight-binding model for
describing the low energy transport  properties of the graphene
bilayer. In this model we consider only the hopping between the
$p_{z}$ orbitals residing on the nearest neighbor sites. In this
picture the two layers are assumed to communicate only through the
hopping among $A_{1}$ and $A_{2}$ sublattices in the Bernal stacking. The
Hamiltonian can be written as:
\begin{eqnarray}\label{Hamilton}
 H=& -&t\sum_{m=1}^{2}\sum_{\langle i,j\rangle}{a_{m,i}^{\dagger} b_{m,j} + h.c.}\\ \nonumber
   &-& t_{\perp}\sum_{i}{a_{1,i}^{\dagger} a_{2,i}+h.c.}\\  \nonumber
   &+& \sum_{m=1}^{2}\sum_{i}\epsilon^{a}_{m,i}a_{m,i}^{\dagger}a_{m,i}+\epsilon^{b}_{m,i}b_{m,i}^{\dagger}b_{m,i},
\end{eqnarray}
in which  $a_{m,i}^\dagger$($a_{m,i}$) creates (annihilates) a
$p_{z}$-electron   at site $(m,i)$ on the sublattice A, with
$m=1,2$, where $m$ is the index of layers and $i$ labels the
sites on each A-sublattice in given layer. Similarly,
$b_{m,i}^\dagger$ and $b_{m,i}$ are the corresponding creation
and annihilation operators in the B-sublattices of the two layers.
Here, $t$ denotes the intra-plane hopping integral between the
nearest neighbors and  $t_{\perp}$ is the inter-plane hopping
amplitude between two layers. Empirical estimates of the hopping
terms are $t \sim 3.16$ eV, and $t_{\perp} \sim 0.39$
eV~\cite{Toy,Misu}. There are  further inter-plane couplings,
where for simplicity we ignore them in our study. Also,
$\epsilon^{a}_{m,i}$ and $\epsilon^{b}_{m,i}$ in the last term
denote the on-site energies at A and B sublattices on each layer,
respectively. To introduce disorder on the model, we choose the
on-site energies  randomly from a uniform distribution in the
interval $[-\frac{W}{2},\frac{W}{2}]$. This is the so called
Anderson model with spatially uncorrelated diagonal disorder  in
which $W$ is a measure for the intensity of disorder. Hereafter,
we assume the unit of energy to be set by $t$.

\section{Kernel Polynomial Method}

Local density of states (LDOS), denoted by  $\rho_{i}(E)$, is
a quantity that measures the contribution of a given lattice site $i$ in
total density of energy states of the lattice in the interval
$[E, E+dE]$, and is defined by the following relation:
\begin{equation}\label{ldos}
{\rho_{i}}(E)= \sum_{n=1}^{N} |\langle
i|E_{n}\rangle|^2\delta(E-E_n),
\end{equation}
in which $|E_{n}\rangle$ is the energy eigenvector corresponding
to energy eigenvalue $E_{n}$. $|\langle i|E_{n}\rangle|^2$ in
Eq.~(\ref{ldos}) is the probability of finding an electron with
energy $E_{n}$ on the site $i$, which in the absence of disorder
is the same for all lattice sites as a result of translational invariance.
However for localized states encountered in disordered systems,
this probability drastically varies over the system.
If the Hamiltonian possesses an extended eigenstate with the
eigenvalue between $E$ and $E+dE$, then all sites are comparably
likely to be present in this state, while for a localized state only  a
limited number of  sites have appreciable probability of being occupied.
Therefore LDOS would be  a suitable quantity to distinguish an
extended state from a localized one. This can be accomplished by
comparing the geometric and arithmetic averagings of LDOS's
at different lattice points. The geometric average of LDOS's
known as typical DOS is defined as:
\begin{equation}\label{typical}
\rho_{\rm typ}(E) =
\exp\left[\frac{1}{K_rK_s}\sum_{r=1}^{K_r}\sum_{i=1}^{K_s}\ln\left(\rho_i^{r}(E)\right)\right],
\end{equation}
where $K_{s}$ is the number of sites in a given realization
that LDOS is calculated and $K_{r}$ is the number of realizations.
On the other hand, the total density of states is obtained by
summing up the partial LDOS of all lattice sites, which amounts
to the following arithmetic average:
\begin{equation}\label{total}
\rho_{av}(E) =
\frac{1}{K_rK_s}\sum_r^{K_r}\sum_i^{K_s}\rho_i^{r}(E)
=\frac{1}{D}\sum_{n=0}^{D-1}\delta(E-E_n),
\end{equation}
where $D$ is the dimension of Hilbert space of the Hamiltonian.
For an extended state $\rho_{\rm typ}(E) \approx \rho_{\rm
av}(E)$, while in the case of localized states $\rho_{\rm
typ}(E)\ll\rho_{\rm av}(E)$.

LDOS  is a site dependent quantity which we use
KPM~\cite{Weisse} to compute it. The basic idea
of KPM is to expand the spectral functions such as,  $\rho_i (E)$, in
terms of orthogonal polynomials of energy $E$.
In principle, one can use any kind of orthogonal polynomials.
In this paper we use  Chebyshev polynomials. Therefore we expand
LDOS as:
\begin{equation}\label{ldos2}
\rho_i(E)=\frac{1}{\pi\sqrt{1-E^2}}\left[\mu_0+2\sum_{n=1}^N \mu_n T_n(E)\right],
\end{equation}
where the coefficient $\mu_{n}$ is given by~\cite{Weisse}:
\begin{equation}\label{coeff1}
\mu_n=\int_{-1}^1 \rho_i(E) T_n(E)dE =\frac{1}{D}\langle
i|T_n(\tilde{H})|i\rangle.
\end{equation}
 $\tilde H$ in the above equation is the rescaled Hamiltonian
which can be obtained by a simple shift and scaling transformation
of the original Hamiltonian to ensure that eigenvalues of
$\tilde H$ lay in the interval $[-1,1]$. A similar procedure for the
expansion of total DOS gives the following moments:
\begin{equation}\label{coeff2}
 \mu'_n = \int_{-1}^1 \rho_{av}(E) T_n(E)
   dE =\frac{1}{D} \mbox{Tr}[T_n(\tilde H)].
\end{equation}
The moments given in Eqs.~(\ref{coeff1}) and (\ref{coeff2}),
can be evaluated employing the recursion relation of Chebyshev
polynomials first discussed by Wang~\cite{Lin1}.
The matrix elements are computed {\em on the fly}, without
saving any matrix, which is a key aspect of KPM. The
summation over the diagonal matrix elements required in the
trace in Eq.~(\ref{coeff2}) can be performed stochastically
which dramatically reduces the computer time.
These points makes it possible to study reasonably large systems
even on a desktop computer~\cite{Weisse}.

When the series expansion of Eq.~(\ref{ldos2}) is used in numerical calculations,
it has to be truncated at a finite order $N$. This truncation leads
to the infamous Gibbs oscillations in the LDOS's.
To relieve this effect, some standard damping factors~\cite{Bulirsch}
(called $g$-factors) have been suggested~\cite{Weisse,Lin1},
by using which Eq.~(\ref{ldos2}) can be evaluated as follows:
\begin{equation}\label{ldos3}
   \rho_i(E)=\frac{1}{\pi\sqrt{1-E^2}}\left[\mu_0 g_0+2\sum_{n=1}^N \mu_n g_n T_n(E)\right].
\end{equation}
In our work, we found that the following positive $g$-factor (Jackson's)
\begin{equation}\label{gfactor}
   g_n=\frac{(N-n+1)\cos(\frac{\pi n}{N+1})+\sin(\frac{\pi n}{N+1})\cot(\frac{\pi}{N+1})}{N+1},
\end{equation}
is suitable to damp the  oscillations  arose
in the calculation of LDOS's~\cite{Weisse}.

\section  {Results and discussions}
\begin{figure}[t] \includegraphics[angle=-90,width=9.cm]{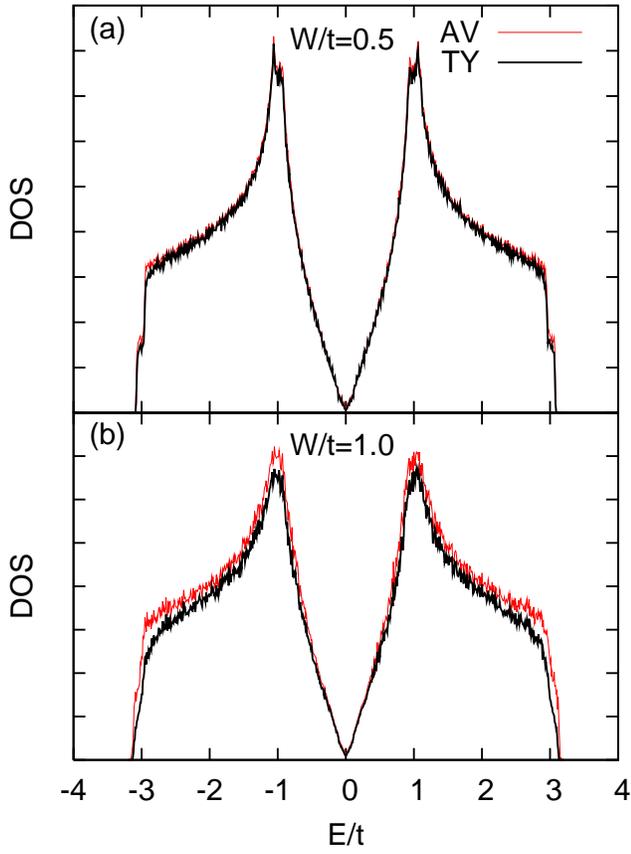}
\caption{ (Color online) Average (red) and typical (black) DOS
for the lattice size  $L=~80\times10^{3}$ with  $K_r\times K_s=
15\times 15$ and $N=4000$ at disorder intensities (a)~$W/t=~0.5$
and (b)~$W/t=~1.0$} \label{weak.fig}
\end{figure}

We operate with the Hamiltonian given in Eq.~(\ref{Hamilton})
with different strengths of disorder $(W)$ to investigate the
Anderson transition on the bilayer graphene and then calculate
LDOS's by means of KPM. We carried out the calculations for the
lattices consisting of $L=20\times 10^{3}$  to $L=80\times
10^{3}$ sites.  In order to obtain reliable results, the order of expansion,
$N$, has to be reapportioned depending on the system size as $N\sim L^2$.
The number of random lattice sites used in averaging the LDOS's
is $K_{s}=15$ for each of the $K_r=15$ different realizations.
In Fig.~\ref{weak.fig}-(a) we compare the total and typical DOS corresponding
to $L=8\times 10^{3}$ sites and $N=4000$ moments for $W/t=0.5$ and
$W/t=1.0$. As can be seen in this figure
$\rho_{typ}$ is nonzero and almost equal to $\rho_{av}$ for the entire
energy bandwidth, indicating  that none of the states are
localized for these strengths of disorder. The four sharp peaks
located  near $E/t\approx\pm 1.0$ in the $W/t=0.5$ plot are the
Van-Hove singularities due to the four saddle points in the band
structure of the clean bilayer graphene. Therefore, this disorder strength
does not appreciably alter the overall spectral aspects of the
clean bilayer graphene. There are also four jumps near
$E/t=\pm 3$, corresponding to the four extrema in energy surfaces.
As can be seen in Fig.~\ref{weak.fig}-(b), for $W/t>1.0$ these singularities tend to smear out.
One has to note that the total bandwidth of clean bilayer graphene is
very close to the value $3t$ of a mono-layer sample, due to the dominant
in-plane energy scale, $t$.
However, as we will see shortly,
when the disorder is introduced, the presence of the second layer
causes the states in the bilayer graphene resist more against
Anderson localization than the mono-layer.

\begin{figure}[t] \includegraphics[angle=-90,width=12.cm]{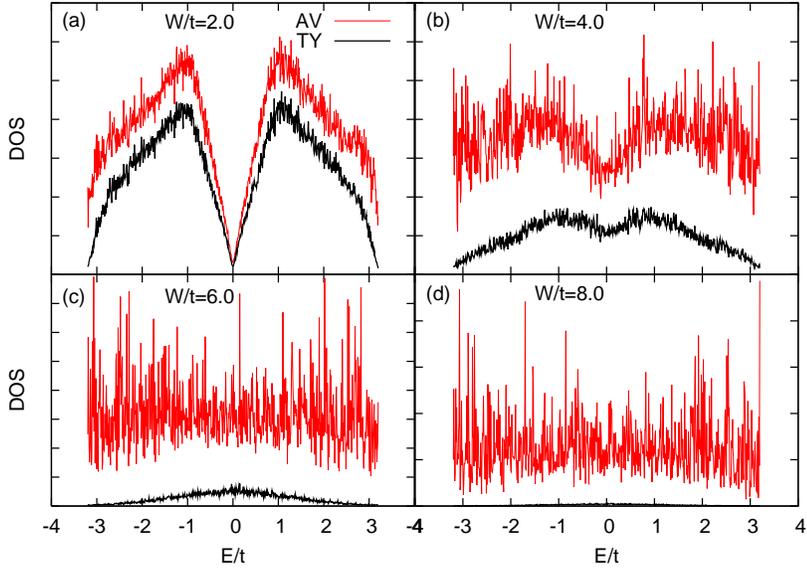}
\caption{ (Color online) Average (red) and typical (black) DOS
for the lattice size  $L=~80\times10^{3}$ with  $K_r\times K_s=
15\times 15$ and $N=4000$ at disorder intensities
(a)~$W/t=~2.0$, (b)~$W/t=~4.0$, (c)~$W/t=~6.0$ and
(d)~$W/t=~8.0$.}\label{strong.fig}
\end{figure}

The results for the larger values of the  disorder strength are
displayed in Fig.~\ref{strong.fig}. As can be seen in this figure,
the average density of states still resembles that of the clean
graphene bilayer up to $W/t=2$. Beyond $W/t\sim 2$, the spectral
properties start to significantly deviate from the clean bilayer.
The typical DOS starts to vanish at two bounds of the energy
spectrum. This behavior is quite similar to the localization
behavior of 3D bands, where a mobility edge sets in, rendering
the states at the tails localized. Further increase in the
disorder strength, eventually localizes the entire spectrum
beyond $W/t\sim 8$. The fact that states at the band edge are
more sensitive to disorder and get localized before those at the
center of the band is in bold contrast to mono-layer
graphene~\cite{amini,Naumis}. This unusual behavior of mono-layer
graphene has been experimentally observed in photoemission
experiment~\cite{Horn}.

\begin{figure}[t]
\includegraphics[angle=-90,width=10.cm,clip=true]{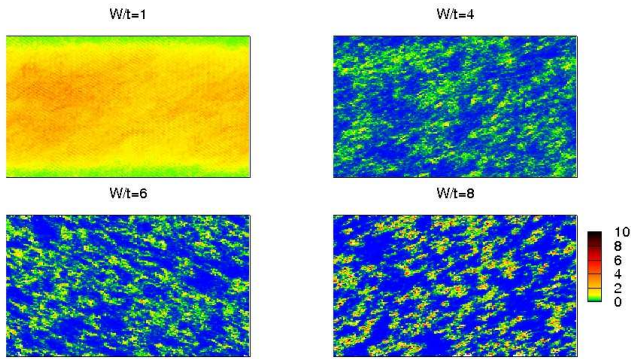}
\caption{ (Color online) LDOS map for various disorder
intensities for $E\in [0.0,0.01]$. In the weak disorder regime,
the electron density is almost uniformly distributed over the
entire lattice. By increasing $W/t$ to $8$, the electron density
becomes confined in disconnected regions of lattice. }
\label{snapshot.fig}
\end{figure}

Fig.~\ref{snapshot.fig} shows intensity plot of LDOS in the
xy-plane of the lattice for different values of disorder near
charge neutrality point, $E=0$. As can be seen, by increasing the
disorder strength, the spatial distribution of the DOS tends to
clump into clusters. At the  disorder strength $W/t=8.0$, these
clusters of non-zero LDOS become entirely disconnected in such a
way that the energy states around $E=0$ would no longer
contribute  to the electric conduction.

\begin{figure}[t]
\includegraphics[angle=-90,width=10.cm,clip=true]{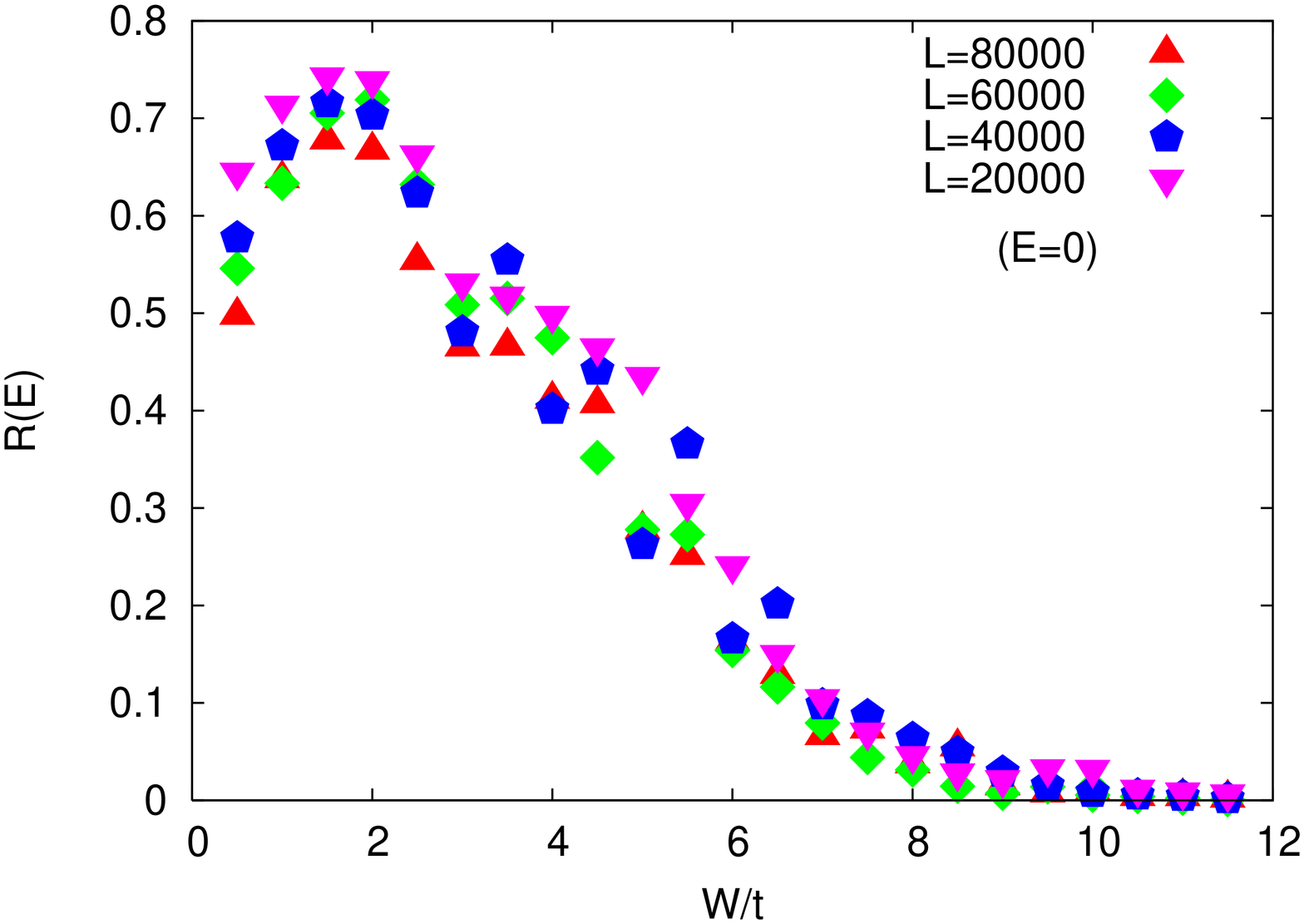}
\caption{ (Color online)  The ratio of typical to average DOS ($R(E)$), computed
at the band center ($E= 0$), versus the strength of disorder for
different lattice sizes $L=2\times 10^4,4\times 10^4,6\times
10^4,8\times 10^4$.} \label{finite-size1.fig}
\end{figure}

\begin{figure}[t]
\includegraphics[angle=-90,width=10.cm,clip=true]{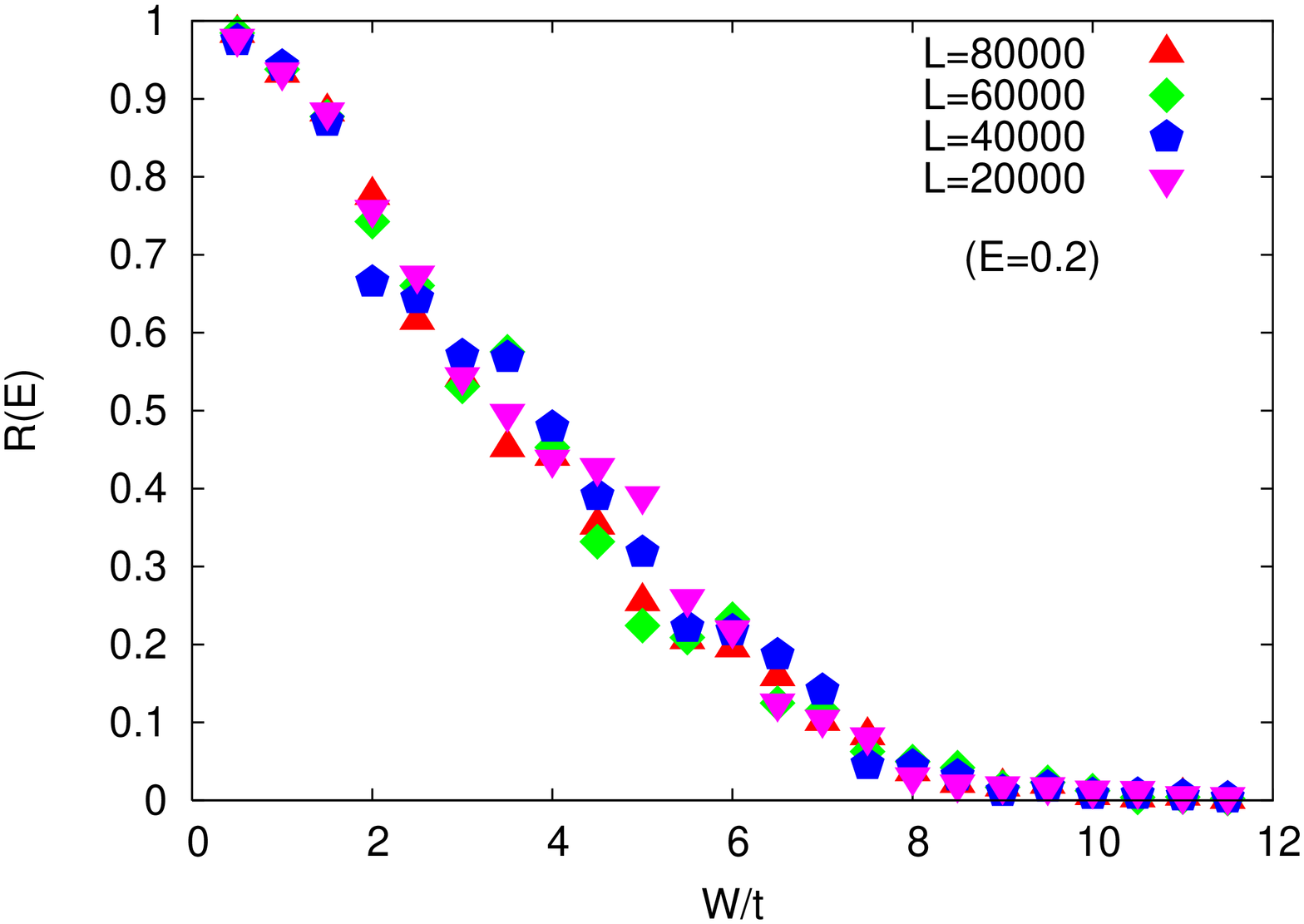}
\caption{ (Color online) The ratio of typical to average DOS ($R(E)$), computed
at $E=0.2$ , versus the strength of disorder for different lattice
sizes $L=2\times 10^4,4\times 10^4,6\times 10^4,8\times 10^4$.}
\label{finite-size2.fig}
\end{figure}

To rule out the possibility of finite size artifacts, we perform a
finite size scaling analysis. Let us define $R(E)$ as the ratio of the
typical to average DOS for a given electronic mode at
energy $E$~\cite{Schubert}:
\begin{equation}\label{order}
   R(E)= \frac{\rho_{typ}(E)}{\rho_{av}(E)}.
\end{equation}
Since always the arithmetic averages of positive numbers are greater
than the geometric average, one always has,
\be
   R(E) \le 1,~~~~~\forall E.
\ee
In the absence of disorder the equality is realized, while turning
the disorder on, reduces $R(E)$ from $1$. We computed this quantity for
different values of disorder strength and different lattice sizes.
The results for two energies $E=0$ and $E=0.2$ are depicted in
Figs.~\ref{finite-size1.fig} and \ref{finite-size2.fig}, respectively.
The lattice sizes are $L=2\times 10^4, 4\times 10^4, 6\times 10^4$ and $8\times 10^4$.
As can be seen in a wide range of disorder strengths, this ratio
tends to converge almost to the same curve with increasing $L$,
which indicates the validity of our results on the infinite lattice size.
We have checked that increasing the order of expansion from $N=4000$
to $N=12000$, does not appreciably change our results.

\section{conclusion}

 In summary,  using KPM method to find the local density of
states; and studying their geometrical averages, we investigated
the effect of uncorrelated  disorder on
the electronic states of the graphene bilayer. The localization
behavior of bilayer graphene is reminiscent of 3D bands, i.e. the
localization starts from the states at the edges of the energy
spectrum. The states near the charge neutrality point $E=0$
remains extended up to very large value of disorder strength
$W/t\sim 8$. Therefore, our finding shows that the
bilayer graphene remains metal in the presence of on-site disorder
up to a critical value of disorder at which undergoes the Anderson
metal-insulator transition. This critical value  is about three
times larger than the critical value that we obtained  for
Anderson transition in single-layer graphene~\cite{amini}.
The inter-layer hopping seems to be responsible for such a
drastic difference between the localization properties of
mono-layer and bilayer graphene. This difference manifests
itself in more than twice a large minimal conductivity of
bilayer graphene compared to mono-layer samples.
The precise understanding of how the (small) inter-layer hopping can
lead to such a remarkable difference in the localization properties
remains an open question which needs further investigations.

\section {Acknowledgment}
This research was supported by the Vice Chancellor for Research
Affairs of the Isfahan University of Technology (IUT). S. A. J was
supported by the National Elite Foundation (NEF) of Iran. This
work was partially supported by Nanotechnology initiative of Iran.
Numerical calculations  of this research were carried in the IUT
Advanced Computational Center.


\end{document}